\documentclass[a4paper, 12pt]{article}
\usepackage{graphicx}
\def \d {{\rm d}}

\begin{document}

\title{\bf On the parameters of the Kerr--NUT--(anti-)de~Sitter space-time} 

\author{J. B. Griffiths$^1$\thanks{E--mail: {\tt J.B.Griffiths(at)Lboro.ac.uk}} \ 
and J. Podolsk\'y$^2$\thanks{E--mail: {\tt podolsky(at)mbox.troja.mff.cuni.cz}}
\\ \\ \small
$^2$Department of Mathematical Sciences, Loughborough University, \\
\small Loughborough,  Leics. LE11 3TU, U.K.\\ 
\small $^1$Institute of Theoretical Physics, Charles University in Prague,\\
\small V Hole\v{s}ovi\v{c}k\'ach 2, 18000 Prague 8, Czech Republic. } 

\date{\today}
\maketitle

\begin{abstract}
\noindent
Different forms of the metric for the Kerr--NUT--(anti-)de~Sitter space-time are being widely used in its extension to higher dimensions. The purpose of this note is to relate the parameters that are being used to the physical parameters (mass, rotation, NUT and cosmological constant) in the basic four dimensional situation. 
\end{abstract}

\bigskip\bigskip

The important family of Kerr--NUT--(anti-)de~Sitter space-times have recently been generalised to arbitrary higher numbers of dimensions \cite{ChLuPo06b}--\cite{HaHoOoYa07}. These space-times are being widely investigated and some of their particular properties have already been found (see \cite{ChGiLuPo05}--\cite{FrKrKu07} and references therein). However, in many of these investigations, the physical significance of the parameters employed have not been clarified -- even in the standard 4-dimensional background. The purpose of this note is simply to identify the parameters of this family of solutions in the 4-dimensional case.

The form of the metric for the Kerr--NUT--(anti-)de~Sitter family of space-times that has been employed reduces in four dimensions to 
  \begin{equation}
 \d s^2= -{Q\over r^2+p^2}(\d\tau-p^2\d\sigma)^2
 +{P\over r^2+p^2}(\d\tau+r^2\d\sigma)^2 
 +{r^2+p^2\over P}\,\d p^2
 +{r^2+p^2\over Q}\,\d r^2 ,
  \label{KNUTAdSmetric}
  \end{equation}
  where $P(p)$ and $Q(r)$ are quartic functions whose coefficients are related in a specific way. For physically significant (black hole-type) solutions, $P$ and $Q$ must have at least two roots. To maintain a Lorentzian signature, it is necessary that ${P>0}$, and so $p$ is constrained to the appropriate range between the roots (poles). The roots of $Q$ correspond to the horizons.

\subsubsection*{First form} 
In~\cite{ChGiLuPo05} and \cite{KubFro07}, the metric functions have been expressed in the form
 \begin{equation} 
 P =k +2np -\epsilon p^2 -{\textstyle{1\over3}}\Lambda p^4, \qquad
 Q =k -2mr +\epsilon r^2 -{\textstyle{1\over3}}\Lambda r^4,
 \label{firstPQeqns}
 \end{equation} 
 where $\Lambda$ is the cosmological constant, and the four parameters $m$, $n$, $k$ and $\epsilon$ are related to the mass of the source, the rotation parameter~$a$ and the NUT parameter~$l$, with one freedom of scaling.

In fact these solutions are members of the Pleba\'nski--Demia\'nski \cite{PleDem76} family, which have recently been described in detail in~\cite{GriPod06}. Taking the vacuum, non-accelerating solutions, the expressions~(\ref{firstPQeqns}) are obtained in the notation of~\cite{GriPod06} by scaling ${\omega=1}$. In this case, $m$ represents the mass of the source, and the other parameters are related to the physical parameters ($a$, $l$, $\Lambda$) as 
 \begin{equation} 
 \epsilon =a_0-(a^2+6l^2){\textstyle{1\over3}}\Lambda, \qquad
 n=l\left[a_0+(a^2-4l^2){\textstyle{1\over3}}\Lambda\right], \qquad
 k=(a^2-l^2)(a_0-l^2\Lambda), 
 \label{epsilonnk} 
 \end{equation} 
 where $a_0>0$ to ensure that $P>0$ between its appropriate roots. It is then convenient to use the remaining scaling freedom to put ${a_0=1}$. This is achieved by putting 
 \begin{equation} \begin{array}{c}
 p=a_0^{1/2}\,\tilde p, \qquad r=a_0^{1/2}\,\tilde r, \qquad
 \tau=a_0^{-1/2}\,\tilde\tau, \qquad \sigma=a_0^{-3/2}\,\tilde\sigma, \\[4pt] 
 k=a_0^{2}\,\tilde k, \qquad n=a_0^{3/2}\,\tilde n, \qquad 
 m=a_0^{3/2}\,\tilde m, \qquad \epsilon=a_0\,\tilde\epsilon, 
 \end{array}
 \label{scaling} 
 \end{equation} 
 and subsequently ignoring the tilde.

With $a_0=1$, the equations (\ref{epsilonnk}) identify the physical meaning of the parameters of~(\ref{firstPQeqns}). Of course, these relations between the parameters of~(\ref{firstPQeqns}) and the ``genuine'' physical parameters are not unique. But they are correct to the extent that the resulting metric reduces to the Kerr--\hbox{(anti-)de~Sitter} metric when ${l=0}$ and the NUT--(anti-)de~Sitter metric when ${a=0}$. With these identifications, the metric functions (\ref{firstPQeqns}) take the forms 
 $$ \begin{array}{l}
 P=\Big[a^2-(p-l)^2\Big]\left[1+(p-l)(p+3l){1\over3}\Lambda\right], \\[8pt]
 Q=a^2-l^2 -2mr +r^2 -
 \Big[3l^2(a^2-l^2)+(a^2+6l^2)r^2+r^4\Big]{\textstyle{1\over3}}\Lambda. \end{array} $$ 
 This clearly identifies the required roots of $P$, and it is appropriate to make the substitutions 
 $$ p=l+a\cos\theta, \qquad \tau=t-{(a+l)^2\over a}\phi, \qquad 
 \sigma=-{\phi\over a}, $$ 
 where ${\theta\in(0,\pi)}$, so that the metric (\ref{KNUTAdSmetric}) becomes 
  \begin{equation}
  \begin{array}{l}
 {\displaystyle \d s^2=
 -{Q\over\varrho^2}\left[\d t- \left(a\sin^2\theta
+4l\sin^2{\textstyle{\theta\over2}} \right)\d\phi \right]^2
   +{\varrho^2\over Q}\,\d r^2 } \\[8pt]
  \hskip8pc {\displaystyle 
  +{\tilde P\sin^2\theta\over\varrho^2} \Big[ a\d t  
  -\Big(r^2+(a+l)^2\Big)\d\phi \Big]^2  
 +{\varrho^2\over\tilde P}\,\d\theta^2 , }
\end{array}
  \label{newKNUTAdSmetric}
  \end{equation}
  where 
 $$ \varrho^2 =r^2+(l+a\cos\theta)^2, \qquad
  \tilde P= 1+{\textstyle{4\over3}}\Lambda al\cos\theta
 +{\textstyle{1\over3}}\Lambda a^2\cos^2\theta. $$ 
 This nicely reduces to the familiar forms in all the particular subcases.

\subsubsection*{Second form} 
In other recent papers, such as \cite{ChLuPo06b}, \cite{ChLuPo07}, \cite{PaKuVaKr07} and \cite{FrKrKu07} the metric functions (in 4D) have been expressed in the alternative form 
 \begin{equation} 
 P=(A^2-p^2)(1+{\textstyle{1\over3}}\Lambda p^2) +2Lp, \qquad 
 Q=(A^2+r^2)(1-{\textstyle{1\over3}}\Lambda r^2) -2Mr,
 \label{secondPQeqns} 
 \end{equation} 
 where $M$, $A$ and $L$ are different constant parameters. When ${L=0}$ and ${p=A\cos\theta}$, this reduces to the Kerr--(anti-)de~Sitter solution in which $M$ is the mass~$m$ of the source and $A$ is the rotation parameter~$a$. However, these parameters do not generally have this identification.

By comparing coefficients in the quartics (\ref{firstPQeqns}) and (\ref{secondPQeqns}), it can be seen that
 $$ M=m, \qquad A^2=k, \qquad L=n, $$ 
 with the additional constraint that $\epsilon=1-{1\over3}\Lambda A^2$. In this case, the metric functions (\ref{secondPQeqns}) have the same number of parameters as physical parameters, so no scaling freedom remains. By inserting the expressions (\ref{epsilonnk}) into this additional constraint, we obtain that 
 $$ a_0= {1+(a^2+3l^2){1\over3}\Lambda \over 1+(a^2-l^2){1\over3}\Lambda} 
 +l^2\Lambda \
 = \ 1+l^2\Lambda{7+(a^2-l^2)\Lambda\over3+(a^2-l^2)\Lambda}. $$ 
 This expression is required to be positive, which it generally is. After inserting this into (\ref{epsilonnk}), the final expressions for the parameters of (\ref{secondPQeqns}) are given by 
 $$ A^2=\left(a^2-l^2\right) \left({1+(a^2+3l^2){1\over3}\Lambda \over 
 1+(a^2-l^2){1\over3}\Lambda}\right), \qquad
 L=l\left[ {1+(a^2+3l^2){1\over3}\Lambda \over 1+(a^2-l^2){1\over3}\Lambda}
 +\left(a^2-l^2\right){\textstyle{1\over3}}\Lambda \right]. $$ 
 This gives the correct Kerr--(anti-)de~Sitter solution when $\l=0$, but the NUT--(anti-) de~Sitter solution requires a negative value for $A^2$. However, as this is just a parameter of (\ref{secondPQeqns}), there is no reason why it should not take negative values when ${a^2<l^2}$. With this identification of parameters, the metric can be transformed to the form (\ref{newKNUTAdSmetric}) after first applying the rescaling~(\ref{scaling}).

It has been shown in~\cite{GriPod05} that the Kerr--NUT family of solutions have to be distinguished into two types. Firstly, there are Kerr-like solutions with a small NUT parameter for $a^2>l^2$. These have a curvature singularity at $r=0$, $\cos\theta=-l/a$ so that $r\in(0,\infty)$. And secondly, there are NUT-like solutions with a small rotation for $a^2<l^2$. These have no curvature singularity so that ${r\in(-\infty,\infty)}$.  The solutions expressed using the notation of (\ref{secondPQeqns}) represent the first type when $A^2>0$ and the second type when $A^2<0$.

\subsubsection*{Acknowledgements}

This work was partly supported by grants from the EPSRC, by GACR 202/06/0041 and the Czech Centre for Theoretical Astrophysics LC06014.

\end{document}